\documentclass[twocolumn,secnumarabic,amssymb, nobibnotes, aps, prd, showpacs]{revtex4}

\usepackage{amsmath,amsthm,amssymb}
\usepackage[dvips]{graphicx,color}
\begin{document}

\title{Solvable Stochastic Dealer Models for Financial Markets}%
\author{Kenta Yamada$^1$}\email[E-mail: ]{yamada@smp.dis.titech.ac.jp}
\author{Hideki Takayasu$^2$}
\author{Takatoshi Ito$^3$}
\author{Misako Takayasu$^1$}
\affiliation{$^1$Department of Computational Intelligence and Systems Science, Interdisciplinary Graduate School of Science and Engineering, Tokyo Institute of Technology, 4259 Nagatsuta-cho, Midori-ku, Yokohama 226-8502, Japan}
\affiliation{$^2$Sony Computer Science Laboratories, 3-14-13 Higashi-Gotanda, Shinagawa-ku, Tokyo 141-0022, Japan}
\affiliation{$^3$Faculty of Economics, The University of Tokyo, 7-3-1 Hongo, Bunkyo-Ku, Tokyo 113-0033, Japan}

\begin{abstract}
We introduce solvable stochastic dealer models, which can reproduce basic empirical laws of financial markets such as the power law of price change. Starting from the simplest model that is almost equivalent to a Poisson random noise generator, the model becomes fairly realistic by adding only two effects, the self-modulation of transaction intervals and a forecasting tendency, which uses a moving average of the latest market price changes. Based on the present microscopic model of markets, we find a quantitative relation with market potential forces, which has recently been discovered in the study of market price modeling based on random walks.

\end{abstract}
\maketitle

\section{Introduction}

Research on financial markets using methods and concepts developed in physics has increased considerably over the last decade. Various kinds of stylized facts or empirical laws of markets have been discovered from high precision market data of gigantic size \cite{econophysics-1}\cite{empirical-laws1}\cite{volatility-1}\cite{volatility-2}\cite{EBS-ohnishi}. The next goal of this econophysics study is to attempt to establish the reasons for these empirical findings. Just as with the Boyle-Charles' macroscopic law which can be derived from a simple microscopic ideal-gas model, we hope to construct a simple microscopic model of a market that can reproduce major empirical findings. By relating macroscopic market behavior to microscopic dealers' actions, we may find a pathway to control the markets, so as to avert bubbles and crashes, which occasionally cause problems in the market.

The study of modeling dealers' action is carried out with so-called agent-based models. This approach is supported not only by economists but also by information scientists and physicists \cite{Lux}\cite{Izumi}\cite{Levy}. Agent-based models can in practice reproduce dealers' actions in the market and they can also reflect empirical laws of markets to some extent. However, agent models generally include a huge number of parameters, and it has proved difficult to understand the relation between the parameters of the model and resulting market behavior.

In order to find relationships between the parameters of dealers' actions and market behavior, we have already introduced a kind of minimal model of an agent-based market which consists of dealers with simple deterministic time evolution rules \cite{dealermodel1}\cite{dealermodel2}\cite{dealermodel3}. With this model, we successfully reproduced most of the basic empirical laws using a minimal number of parameters, and found that there are only three important effects needed to reproduce the empirical laws. The first effect is the compromise pricing of both buyers and sellers, who tend to allow the particular transaction price they have in mind to approach the current market price in order to make a deal. From this effect, transactions occur spontaneously in the market and the price rises and falls almost randomly. The second effect is the self-modulation of transaction intervals, that is, the rate of a dealer's clock depends on the latest moving average value of transaction intervals. When market activity becomes high, dealers accelerate their transaction rates, and by this effect we can reproduce empirical statistical properties of transaction intervals which deviate from a simple Poisson process. The third is the trend-follow effect, that is, dealers forecast upcoming prices using the latest market trend which is defined by a moving average of price changes. This forecasting effect makes the price change distribution follow a power law quite similar to that of the real market.

In this paper we first introduce a stochastic version of the dealer model which is even simpler than the above (deterministic) model. In the case of the deterministic dealer model we needed at least three dealers to reproduce market properties; however, in the present stochastic model we require only two. The advantages of this stochastic model are not only its simplicity but also its solvability by analytical calculation. In the usual agent-based approaches intensive numerical simulation is the only way to obtain results; in such cases exact or strict results are rarely obtained. Based on this stochastic dealer model and its variants we can derive the major empirical results mentioned above, that have already been obtained by simulation of the deterministic dealer model by theoretical analysis.

Apart from agent-based modeling, the standard way to model markets is by utilizing random walks. It is now widely known that Bachelier introduced a random walk model for market prices five years earlier than Einstein's random walk model of Brownian motion\cite{Bachelier}\cite{Einstein}. Work on portfolio theory, option price formulation\cite{Black-Scholes} and the ARCH and GARCH models\cite{ARCH}\cite{GARCH}, which has led to Nobel prizes for their developers, are all based on random walk models.

Recently, one of the authors (M.T.) has introduced a new type of extended random walk model of the market, the so-called PUCK model, in which a random walker moves according to a deforming potential force, the center point of which is given by the moving average of the random walker's traces. By using this generalization, all major empirical laws can be established; moreover, dynamical behaviors such as bubbles, crashes and inflations can also be described as following from special cases of the market potential force\cite{PUCK1}\cite{PUCK2}. The ARCH model can also be derived as a special limiting case of this extended random walk model\cite{PUCK-ARCH}. 

Considering the wide applicability of the PUCK model has led to an open question concerning the origin of market potential forces. This question has been partially answered by using the deterministic dealer model \cite{Potential-DM}. Here, we are able to provide quantitative answers by using the stochastic dealer model.

In the next section we introduce our stochastic dealer models step by step in sequential subsections.  The third section is devoted to the relationship with the PUCK model, in which we will see how dealers' actions produce a market's potential force in a quantitative discussion. The final section contains a summary.

\section{The stochastic dealer model}
In this section we introduce three stochastic dealer models, Model-1, Model-2 and Model-3. Model-1 is the simplest market model in which the framework of the stochastic dealer model is introduced. Then, we note two empirical properties which Model-1 cannot reflect. In Model-2 and Model-3 we introduce two additional effects respectively to deal with these difficulties. After combining these revisions, the stochastic dealer model fully reflects all major empirical laws of markets.

\subsection{Model-1}\label{sec:Model-1}
\begin{figure}[htbp]
  \begin{center}
    \includegraphics[width=80mm]{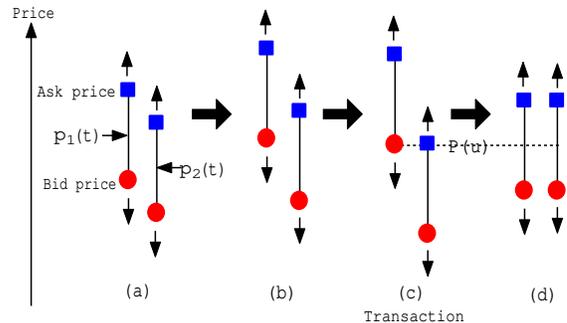}
  \end{center}
  \caption{Time evolution of the dealer model. Squares and circles denote ask and bid prices, respectively. The $i$-th dealer's mid-price is denoted by $p_i(t)$. (a): In this situation no transaction occurs. (b): The dealers' prices follow random walks. (c): When the distance between $p_1(t)$ and $p_2(t)$ is greater than or equal to $L$, a transaction occurs and the market price is defined by the averaged price of the two mid-prices. (d): After this transaction both dealers' mid-prices move to the market price. These processes are repeated. }
\label{fig:transaction2.eps}
\end{figure}

Firstly, we assume an artificial market consisting of only two dealers who are offering both buying and selling prices. The buying price, or bid price, is the current maximum price at which the dealer wants to buy. The selling price, or ask price, is the minimum price at which he will sell. For each dealer the ask price is always higher than the bid price, because they want some margin, and the difference between these prices is called the spread, which is assumed to be a constant, $L$, in this model. We define the $i$-th dealer's mid-price at time $t$, $p_i(t)$, as the average of his bid and ask prices. When $|p_1(t)- p_2(t)|$ is less than $L$, these dealers do not transact as their transaction conditions are not fulfilled (FIG.\ref{fig:transaction2.eps}a). In such a case dealers are assumed to change their prices randomly according to the following rule.
\begin{eqnarray}
p_i(t+\Delta t)&=&p_i(t)+c f_i(t)\quad i\in 1,2,\label{eq:model-1}\\
f_i(t)&=&
\begin{cases}
+\Delta p\quad\text{(prob. 1/2)}\nonumber\\
-\Delta p\quad\text{(prob. 1/2)}
\end{cases}.
\end{eqnarray}
Here, $f_i(t)$ is a random noise for the $i$-th dealer, and $c$ is a constant parameter. Then, the distance between $p_1(t)$ and $p_2(t)$ is checked. If it is greater than or equal to $L$, then one dealer's bid price is higher than the other's ask price, and a transaction occurs (FIG.\ref{fig:transaction2.eps}c). In such a case a unit volume deal is assumed to be made, with the market price given by the averaged price of the two dealers' mid-prices. After this transaction their mid-prices are assumed to shift to the market price. These processes are repeated again and again and the time proceeds in unit of $\Delta t$.

It should be noted that there is a possibility that this model produces a negative value for market price. In such a case the step width of the price change, $\Delta p$, should depend on the market price, such that the value of $\Delta p$ is proportional to the market price to avoid crossing the origin. Here, we pay attention only to the case that the price fluctuation level is much smaller than the market price and for simplicity, we assume that $\Delta p$ is constant.  

For convenience of analysis we define another unit of time called the tick time, denoted by $n$, which takes an integer value incremented at each occurrence of a transaction. Accordingly, $P(n)$ denotes the market price at tick time $n$, and the $n$-th transaction interval, $I(n)$, is defined by the time difference between the $n-1$-th and $n$-th transactions.

\begin{figure}[htbp]
  \begin{center}
    \includegraphics[width=80mm]{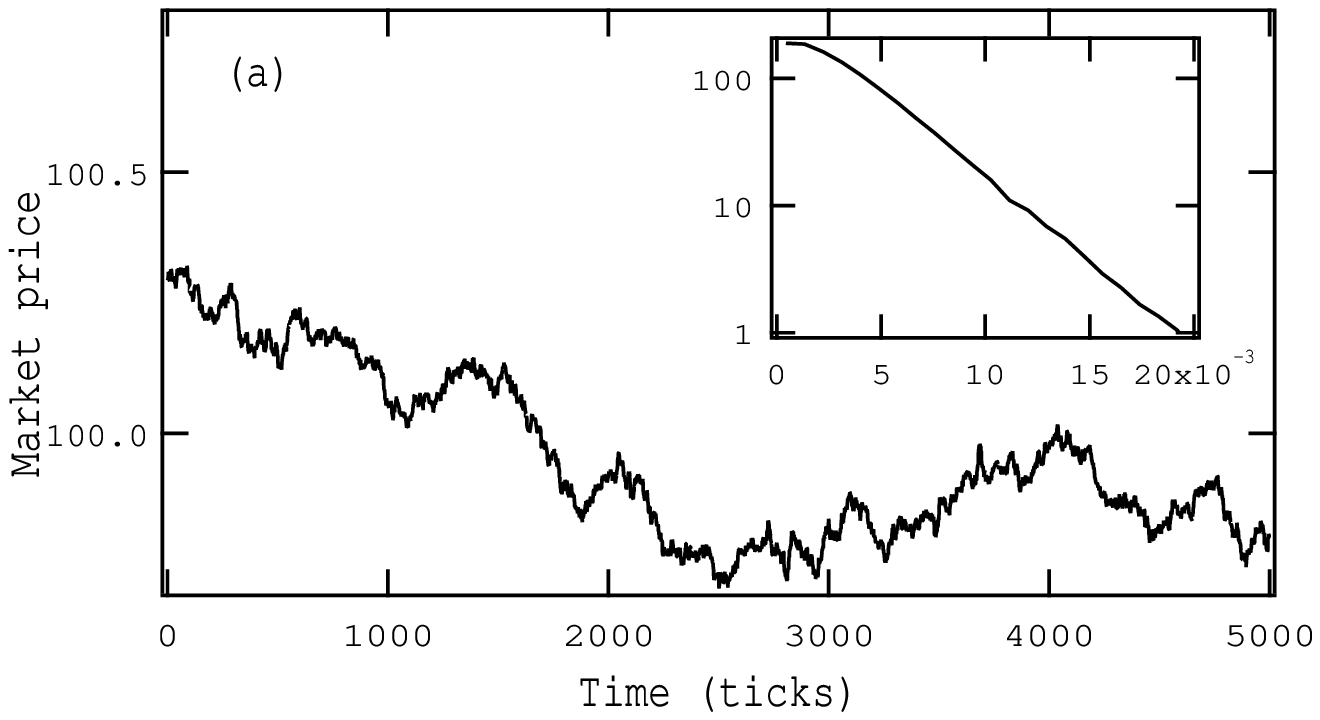}
    \includegraphics[width=80mm]{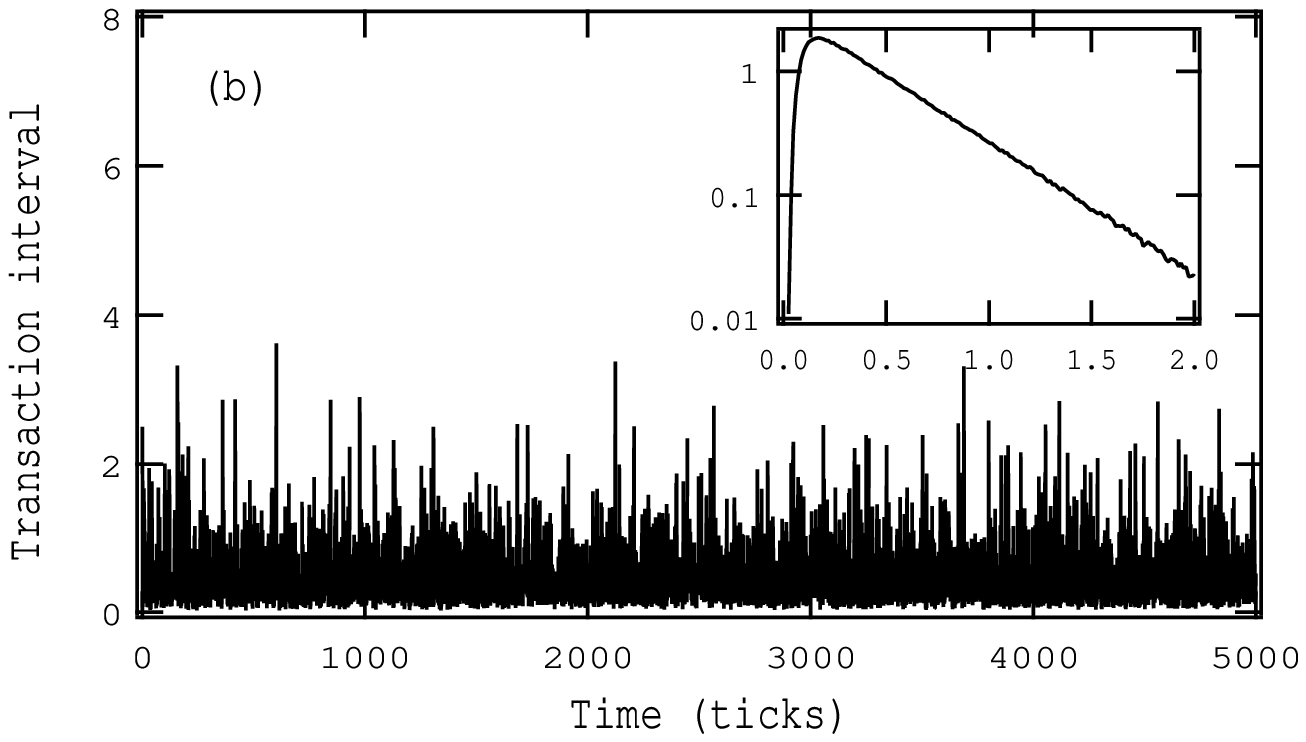}
  \end{center}
  \caption{Examples of time series of market prices (a) and transaction intervals (b). Sub-windows of these figures show the probability densities of market price changes and transaction intervals on a semi-log scale. The parameters for this simulation are as follows: $L=0.01$, $c=0.01$, $\Delta p=0.01$ and $\Delta t=\Delta p\cdot \Delta p$.}
\label{fig:ts-pI}
\end{figure}

In FIG.\ref{fig:ts-pI}, we plot an example of resulting market prices and corresponding transaction intervals. In the sub-windows of these figures we also plot the probability density function of price changes $|P(n)-P(n-1)|$ and transaction intervals $I(n)$ both on a semi-log scale. It is clear that the tail parts of both of these distributions are well characterized by exponential laws. As for the intervals, this result implies that the occurrence of transactions of this model is approximated by a Poisson process. It is interesting that the price change distribution of this simplest model follows an exponential distribution except around $\Delta P=0$, instead of Gaussian distribution.

\begin{figure}[htbp]
  \begin{center}
    \includegraphics[width=80mm]{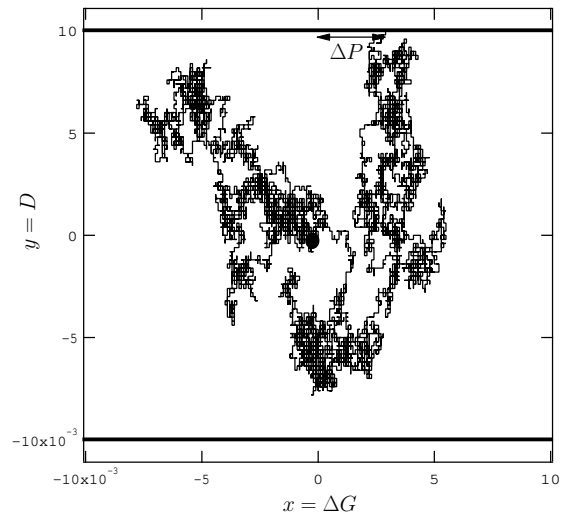}
  \end{center}
  \caption{Random walk in the $\Delta G-D$ plane. The particle starts from the origin and continues a random walk until it touches one of the horizontal walls, meaning that a market transaction occurs. The transaction interval is given by the survival time of this random walker, and the market price change $\Delta P$ is given by the distance along the $\Delta G$ axis from the origin to the position of the particle on the absorbing wall. }
  \label{fig:FootPrint_D-DG_Model-1.eps}
\end{figure}

We can explain the functional form of the tails of these distributions as follows. We define the difference of the dealers' prices by $D(t)=p_1(t)-p_2(t)$, then the condition for the occurrence of a transaction is given by $|D(t)|\ge L$, and we also define the mass center by the average of the two mid-prices, $G(t)=\left\{p_1(t)+p_2(t)\right\}/2$. As the mass center at the time of transaction gives the market price, market price statistics can be calculated from the information about $G(t)$ for times $|D(t)|\ge L$. Namely, $\Delta P$ is given by $\Delta G$ which is defined by $\Delta G=G(t)-G(t^{\prime})$ , where $t^{\prime}$ is the previous transaction time. These two variables define a two dimensional random walk with absorbing walls at $D(t) = L$ and $D(t)= -L$ as shown in FIG.\ref{fig:FootPrint_D-DG_Model-1.eps}. The stochastic dynamics is described by the following set of equations (\ref{eq:2D-brownian_M1}). 

\begin{subequations}
\label{eq:2D-brownian_M1} 
\begin{eqnarray}
D(t+\Delta t)&=&D(t)+
\begin{cases}
+2c\Delta p\quad \text{(prob. 1/4)}\\
\pm 0\quad\quad\ \ \> \text{(prob. 1/2)},\\
-2c\Delta p\quad \text{(prob. 1/4)}
\end{cases}\label{eq:2D-brownian_M1-1}
\end{eqnarray}
\begin{eqnarray}
\Delta G(t+\Delta t)&=&\Delta G(t)+
\begin{cases}
+c\Delta p \quad \text{(prob. 1/4)}\\
\pm 0\quad\quad\  \text{(prob. 1/2)}.\\
-c\Delta p\quad \text{(prob. 1/4)}
\end{cases}\label{eq:2D-brownian_M1-2}
\end{eqnarray}
\end{subequations}

When the random walker reaches one of the absorbing walls a transaction occurs, and by the transaction rule of Model-1 that the prices of the two dealers are then set to the market price, the random walker goes to the origin, and a new random walk begins. In this 2-dimensional formulation the transaction interval, $I(n)$, is given by the survival time, that is, the time the random walker starting from the origin takes to reach one of the absorbing walls. Similarly, the market price change, $\Delta P(n)$, is given by the random walker's location on the $ G$ axis. 

In Fig.3 an example of a random walk is shown for better understanding of this mapping of Model-1 to a 2-dimensional random walk. Here, the horizontal axis is the $x$-axis and the vertical one is the $y$-axis. For theoretical analysis we consider a continuum limit such that the mesh sizes of space and time go to zero. Under the condition  $\Delta t=(\Delta x)^2=(\Delta y)^2$, we find that the probability density $u(x,y,t)$  of the particle  in the $(x,y)$ plane at time $t$ is described by the following diffusion equation. 
\begin{equation}
\frac{\partial u(x,y,t)}{\partial t}=c^2\left (\frac{1}{4}\frac{\partial ^2 u}{\partial x^2}+\frac{\partial ^2u}{\partial y^2}\right ).\label{eq:dif.eq}
\end{equation}
\begin{eqnarray}
\begin{cases}
u(x,y,0)=\delta(x,y-L)\qquad\  \text{:Initial condition}\\
u(x,0,t)=u(x,2L,t)=0\quad \text{:Boundary condition}\label{eq:boundary-initial}
\end{cases}
\end{eqnarray}
Here, $c^2$ is equivalent to the diffusion coefficient of this 2-dimensional random walk. The initial condition is the delta function, and the boundary condition is given by the absorbing walls on the $y$ axis, while there is no boundary in the $x$ direction. This diffusion equation is solved exactly as follows,
\begin{equation}
u(x,y,t)=\frac{1}{cL\sqrt{\pi t}}e^{-\frac{x^2}{c^2 t}}\sum_{n=1}^{\infty}\sin \frac{n\pi}{2}\sin P_ny\cdot e^{-c^2 P_{n}^2t}.\label{eq:solution_u}
\end{equation}
Here, $P_n=\frac{n\pi}{2L}$. We obtain the distributions of transaction intervals $Q_1(I)$ and price changes $Q_2(|\Delta P|)$ by calculating distributions of survival times and absorbed points from (\ref{eq:solution_u}).

\begin{subequations}
\label{eq:pdf-I-ADP}
\label{eq:Q1Q2} 
\begin{eqnarray}
Q_1(I)&=&\frac{4}{\pi}\sum_{n=1}^{\infty}\frac{(-1)^{n+1}}{(2n-1)}c^2P_{2n-1}^2e^{-c^2 P_{2n-1}^2 I},\label{eq:pdf-I}\\
Q_2(|\Delta P|)&=&\frac{4}{L}\sum_{n=1}^{\infty}(-1)^{n+1}e^{-\frac{(2n-1)\pi}{L}|\Delta P|}\label{eq:DP-pdf}.
\label{eq:pdf-ADP}
\end{eqnarray}
\end{subequations}
In the case of large values of $I$ and $|\Delta P|$ in Eq.(\ref{eq:Q1Q2}), these summations are dominated by the term of $n=1$. So, the functions $I$ and $|\Delta P|$ can be approximated as 
\begin{subequations}
\label{allequations} 
\begin{eqnarray}
Q_1(I)&\propto & \displaystyle{e^{-\left ( \frac{c\pi}{2L}\right )^{2}I}},\label{eq:I-pdf2}\\
Q_2(|\Delta P|)&\propto& e^{-\frac{\pi}{L}|\Delta P|}\label{eq:DP-pdf2}.
\end{eqnarray}
\end{subequations}

From these results we can derive the exponential laws of interval distributions and the price change distributions already seen in FIG.\ref{fig:ts-pI}. It is confirmed that the theoretical values of the decay constants $(2L/c \pi )^2$ and $L/ \pi$, fit well with the numerical results.

Higher order moments of the distributions of transaction intervals and price changes are also obtained exactly from Eq.(\ref{eq:pdf-I-ADP}). The $k$-th moments of $<Q_1^k(I)>$ and $<Q_2^k(|\Delta P|)>$ are calculated as follows. 
\begin{subequations}
\label{allequations} 
\begin{eqnarray}
<Q_1^k(I)>&=&\frac{L^{2k}\Gamma(k+1)}{c^{2k}\Gamma(2k+1)}E_k,\label{eq:moment-I}\\
<Q_2^k(|\Delta P|)>&=&\frac{4L^k\Gamma(k+1)}{\pi^{k+1}}\beta(k).
\label{eq:moment-ADP}
\end{eqnarray}
\end{subequations}
Here $\Gamma(x)$ is the gamma function and {$E_k$} are the Euler numbers appearing in the expansion of $\displaystyle{\sec x=\sum_{k=0}^{\infty}\frac{E_kx^{2k}}{(2k)!}}$; $E_0=1,E_1=1,E_2=5,E_3=61,\cdots$. $\beta(k)$ is the Dirichlet beta function defined as $\displaystyle{\beta(k)=\sum_{n=0}^{\infty}\frac{(-1)^n}{(2n+1)^k}}$. Using this result we can calculate means and variances of transaction intervals and volatilities with results as shown in TABLE \ref{tb:ave-vari}.

\begin{table}[htbp]
 \caption{Exact solutions for the means and variances of transaction intervals and the absolute value of price changes in Model-1. $K$ is Catalan's constant; $K=\beta(2)=\displaystyle \frac{1}{2}\int_{0}^{\frac{\pi}{2}}\frac{\theta}{\sin \theta}d\theta$ .}
 \label{tb:ave-vari}
 \begin{center}
  \begin{tabular}{ccc}
    \hline
    \hline
       & $\qquad$ Average$\qquad$  & $\qquad$Variance$\qquad$   \\
    \hline
   Interval ($I$)    & $L^2/2c^2$   & $L^4/6c^4$   \\
   Price change ($|\Delta P|$)   &  $4KL/\pi^2$  & $(1/4-16K^2/\pi^4)L^2$  \\
    \hline
    \hline
  \end{tabular}
 \end{center}
\end{table}

\subsection{Model-2}
\begin{figure}[htbp]
  \begin{center}
    \includegraphics[width=80mm]{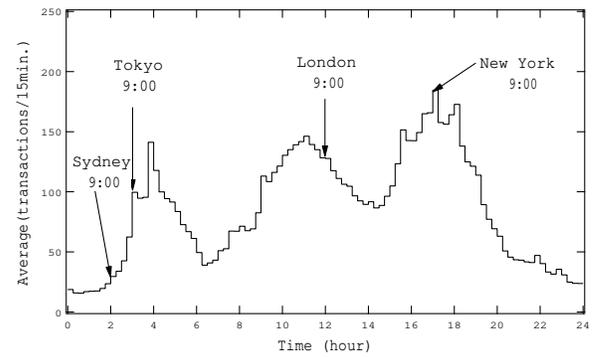}
  \end{center}
  \caption{Diurnal pattern of transactions in the Dollar-Yen market. The vertical axis depicts the mean number of transactions per quarter hour.}
\label{fig:aveNT_15min_YDT_00-05.eps}
\end{figure}

In this subsection we focus on the statistical differences between transaction intervals of real markets and those of Model-1. In real dollar-yen exchange market data provided by EBS for six years from 2000 to 2005, we find that the transaction intervals exhibit a circadian pattern even for Foreign Exchange markets which are open continuously as shown in FIG.\ref{fig:aveNT_15min_YDT_00-05.eps} for the Dollar-Yen market. As can be clearly seen, large numbers of transactions occur during office hours of  Tokyo, London and New York and the transaction density is least a little before the open of the Tokyo offices. 

\begin{figure}[htbp]
 \begin{center}
    \includegraphics[width=80mm]{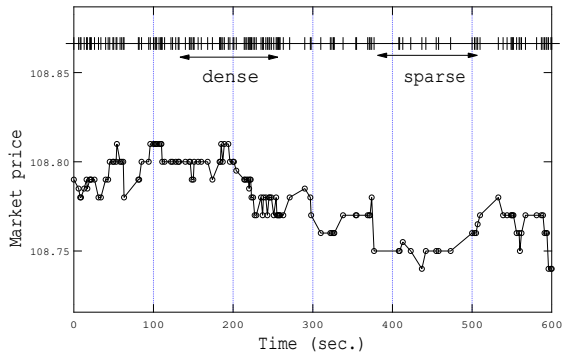}
  \end{center}
  \caption{Dollar-Yen rates for ten minutes (lower curve), and transaction intervals (upper lines). There are dense and sparse periods for the occurrences of transactions. }
\label{fig:Price_20050616_2.eps}
\end{figure}

In addition to this 24-hour pattern, there are fluctuations with much shorter time scales as typically shown in FIG.\ref{fig:Price_20050616_2.eps}. In this figure transaction events are shown by bars at the top and corresponding Dollar-Yen rates are plotted in the lower section. Here the window size is ten minutes and we can find places where bars tend to cluster, marked as "dense", and others where bars are "sparse". The distribution of these intervals is clearly seen not to correspond to the simple theoretical model of a Poisson process. 

The clustering properties of transactions are known to be well modeled by a self-modulation process introduced by the authors \cite{transaction interval}, which is described as follows.
\begin{equation}
x(n+1)=e(n)\cdot<x(n)>_{\tau}+f(n).\label{eq:smp-I}
\end{equation}
where $\tau$ is the time scale of self-modulation, and
\begin{equation}
<x(n)>_\tau=\displaystyle{\frac{1}{N}\sum_{k=0}^{N-1}x(n-k)},
\end{equation}
$e(n)$ and $f(n)$ are independent noises, and $N$ is the number of transactions occurring within $\tau$ seconds. This process is a modified Poisson process whose mean value is given by the moving average of transaction intervals over the past $\tau$ seconds. As a result, there is a greater tendency to cluster and the so-called 1/f fluctuation is realized in general. 

Model-2 is designed to satisfy the real interval property by applying the self-modulation process. We estimate the distribution of $e(n)$ from real data by using the following relation for the transaction intervals, $I(n)$. 

\begin{equation}
e(n)=\frac{I(n)}{<I(n)>_\tau}.
\end{equation}
Here the typical value of $\tau$ is 150 seconds. It is confirmed from Dollar-Yen rate data that the distribution of $e(n)$ follows an exponential distribution in general with mean value of unity. This exponential distribution is favorable for our model construction as Model-1 automatically produces the exponential interval distribution. As we can control the speed of transaction intervals by controlling the speed of diffusion, we obtain a revised model, Model-2, by modifying the constant parameter $c$ in Eq.(\ref{eq:model-1}), which is directly related to the diffusion coefficient, making it a time-dependent parameter $c(n)$ as follows.
 
\begin{eqnarray}
p_i(t+\Delta t)&=&p_i(t)+c(n) f_i(t)\quad i\in 1,2,\label{eq:model-2}\\
f_i(t)&=&
\begin{cases}
+\Delta p\quad\text{(prob. 1/2)}\nonumber\\
-\Delta p\quad\text{(prob. 1/2)}
\end{cases}.
\end{eqnarray}
where 
\begin{equation}
c(n)=\sqrt{\frac{<I>_{c=1}}{<I>_\tau}}\label{eq:c(t)}.
\end{equation}
In Eq.(\ref{eq:c(t)}), $<I>_{c=1}$ is a mean of transaction intervals shown in TABLE \ref{tb:ave-vari} in the case of $c=1$ and $<I>_{c=1}=L^2/2$. $<I>_\tau$ is the moving average of transaction intervals averaged over the latest $\tau$ seconds defined as $\displaystyle{<I>_{\tau}=\frac{1}{N}\sum_{k=0}^{N-1}I(n-k)}$. In this equation, $I(n-k)$ is the transaction interval that is the $k$-th tick earlier than the $n$-th tick. $N$ is the number of transactions within $\tau$ seconds from time $n$. If $I(n)>\tau$, we set $<I>_{\tau}=I(n)$. It is known from Eq.(\ref{eq:c(t)}) that for larger $<I>_{c=1}$ over $<I>_{\tau}$, the value of $c(n)$ is larger, that is, dealers tend to make larger changes their prices to effect more rapid transactions when transaction intervals become shorter in the market. By this effect, Eq.(\ref{eq:2D-brownian_M1}) of Model-1 is modified to
\begin{subequations}
\label{eq:2D-brownian_M2} 
\begin{eqnarray}
D(t+\Delta t)&=&D(t)+
\begin{cases}
+2c(n)\Delta p\; \text{(prob. 1/4)}\\
\pm 0\quad\qquad\; \text{(prob. 1/2),}\\
-2c(n)\Delta p\; \text{(prob. 1/4)}
\end{cases}\label{eq:2D-brownian_M2-1}
\end{eqnarray}
\begin{eqnarray}
\Delta G(t+\Delta t)&=&\Delta G(t)+
\begin{cases}
+c(n)\Delta p \>\  \text{(prob. 1/4)}\\
\pm 0\quad\quad\quad\  \text{(prob. 1/2)}.\ \ \ \\
-c(n)\Delta p\;\  \text{(prob. 1/4)}
\end{cases}\label{eq:2D-brownian_M2-2}
\end{eqnarray}
\end{subequations}

\begin{figure}[htbp]
  \begin{center}
    \includegraphics[width=80mm]{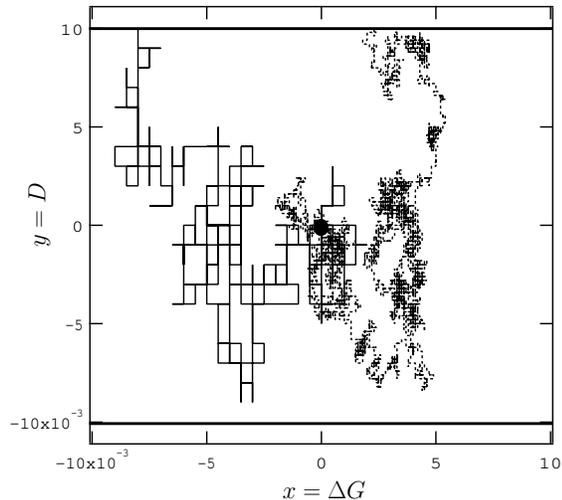}
  \end{center}
  \caption{Examples of random walk traces arising from Model-2. Compared with  FIG.\ref{fig:FootPrint_D-DG_Model-1.eps} for Model-1, the step size depends on past transaction intervals.}
  \label{fig:FootPrint_D-DG_Model-2.eps}
\end{figure}

\begin{figure}[htbp]
  \begin{center}
    \includegraphics[width=80mm]{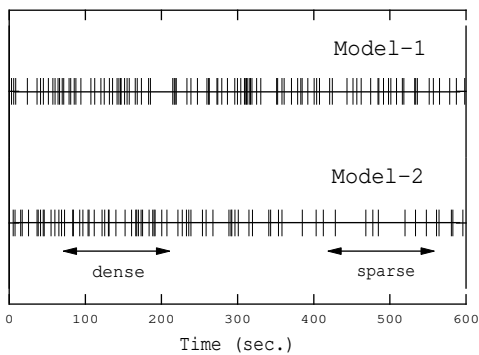}
  \end{center}
  \caption{Examples of transaction occurrences. The upper sequence is produced by a simulation of Model-1 and the lower one by Model-2. }
\label{fig:s-I-model1-model2.eps}
\end{figure}

Examples of random walk traces are shown in FIG.\ref{fig:FootPrint_D-DG_Model-2.eps}. As known from this figure the initial condition and the boundary conditions are the same; however, the step size changes for each random walk following the self-modulation formulation. By this effect the transaction intervals tend to form clusters as shown in FIG.\ref{fig:s-I-model1-model2.eps}.


\begin{figure}[htbp]
  \begin{center}
    \includegraphics[width=80mm]{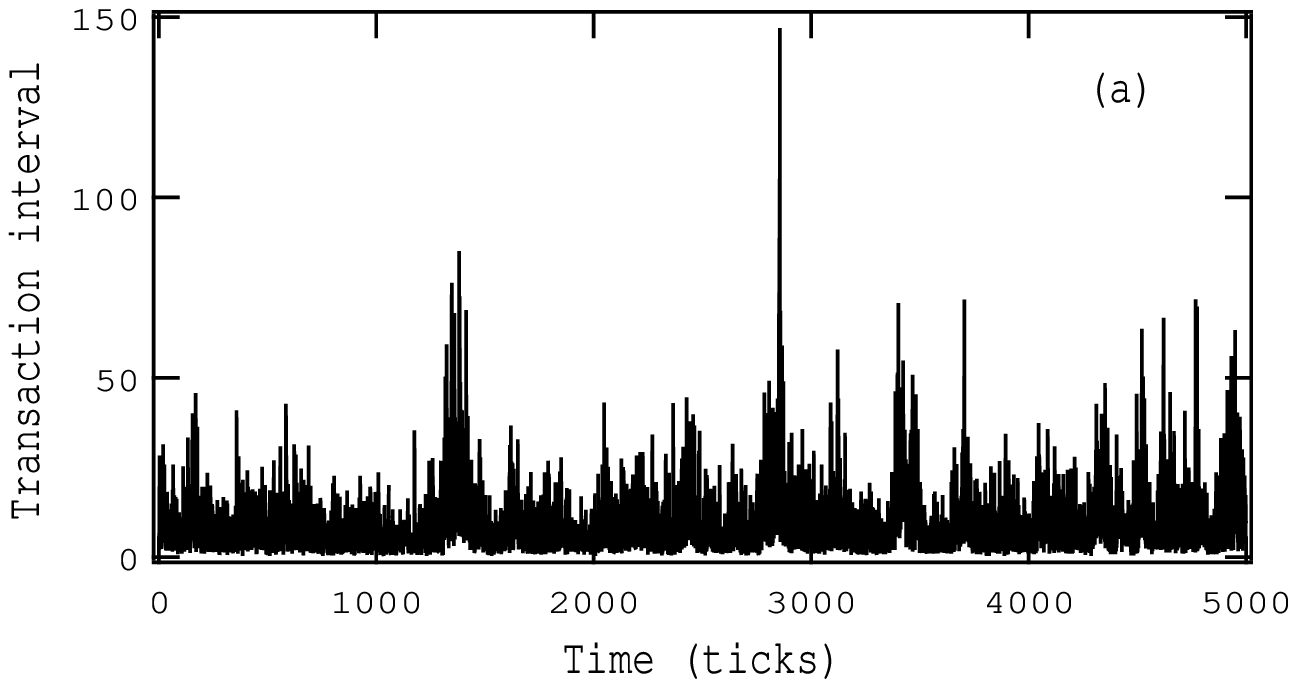}
    \includegraphics[width=60mm]{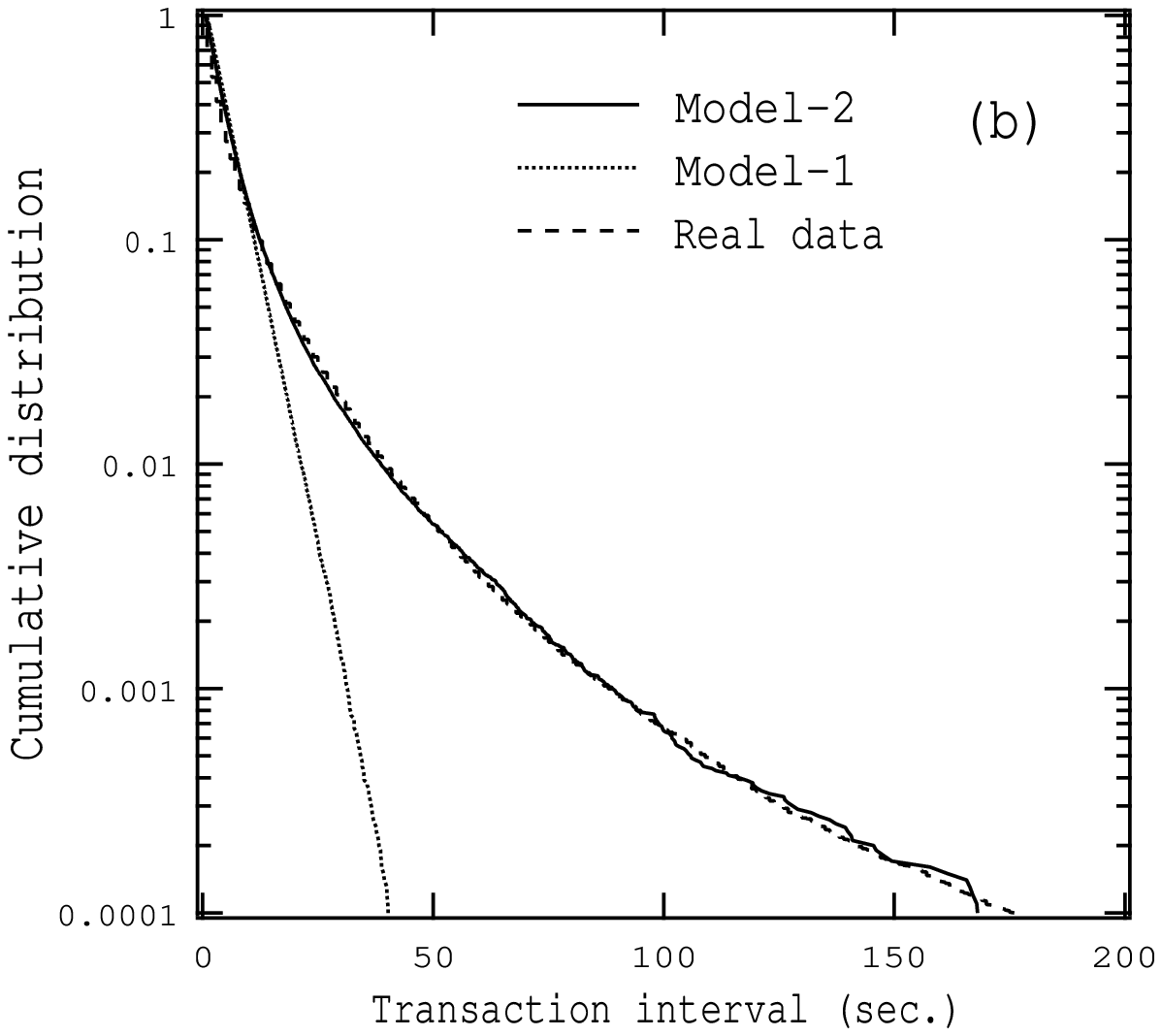}
  \end{center}
  \caption{Comparison of the distributions of transaction intervals. (a) An example of a time series of intervals produced by Model-2. (b) Cumulative distributions of transaction intervals on a semi-log scale. Actual data plotted by dashed line during 9:00-10:00, New York time. The distribution of Model-2 plotted by the solid line has a longer tail than the exponential distribution produced by Model-1 in the dotted line. The distribution of real intervals represented by the dashed line is close to that of Model-2 with the parameters $L=0.01$, $\tau = 150$, $\Delta p=0.01$ and $\Delta t=\Delta p\cdot \Delta p$.}
\label{fig:CDF_I_Model2.eps}
\end{figure}

In order to make the interval distribution fit well with that of the real Dollar-Yen market we introduce two thresholds for the value of $<I>_{\tau}$. When $<I>_{\tau}<3$, we set $<I>_{\tau}=3$, and when $<I>_{\tau}>50$, we set $<I>_{\tau}=50$. These restrictions are needed to prevent intervals from converging to zero, or from diverging to infinity. With these minor revisions an example time sequence produced by Model-2 is plotted together with one produced by Model-1 in FIG.\ref{fig:s-I-model1-model2.eps}. Comparing these two sequences, we observe that Model-2 can reproduce the clustering property quite well. Moreover, the lower interval sequence looks similar to the real sequence shown in Fig.\ref{fig:Price_20050616_2.eps}. Actually, the distribution of transaction intervals arising from Model-2 is now very close to that of the actual interval distribution as shown in FIG.\ref{fig:CDF_I_Model2.eps}.

\subsection{Model-3}\label{sec:model-3}
In this subsection we shift our attention from transaction intervals to price changes. We know that the price change distribution of Model-1 is characterized by an exponential distribution while that of the real market is often characterized by a power law. It has been established that such power law distributions can be derived by introducing the effect of trend-following prediction\cite{dealermodel2}. This effect can be introduced to our stochastic dealer model by simply adding a further term, $d<\Delta P>_M\Delta t$, which is defined as follows.
 
\begin{eqnarray}
p_i(t+\Delta t)&=&p_i(t)+d<\Delta P>_M\Delta t+cf_i(t),\label{eq:model3-1}\\
f_i(t)&=&
\begin{cases}
+\Delta p\quad\text{(prob. 1/2)}\\
-\Delta p\quad\text{(prob. 1/2)}
\end{cases}
i\in 1,2,\nonumber
\end{eqnarray}
where 
\begin{equation}
<\Delta P>_M=\frac{2}{M(M+1)}\sum_{k=0}^{M-1}(M-k)\Delta P(n-k).\label{eq:WMA-1}
\end{equation}
Here $\Delta P(n)=P(n)-P(n-1)$ is the price change at the $n$-th tick. The new term, $<\Delta P>_M$, is a kind of moving average of price changes for $M$ ticks with weights that decay linearly. The parameter $d$ in (\ref{eq:model3-1}) is an important parameter that governs the dealers' strategy. A dealer with positive $d$ is a trend-follower who predicts upcoming market prices proportional to the latest price slope. On the other hand, a dealer with a negative $d$ is called a contrarian who forecasts that upcoming market prices will go against the trend and that the present market price is close to a local maximum or minimum. 

\begin{figure}[htbp]
  \begin{center}
    \includegraphics[width=80mm]{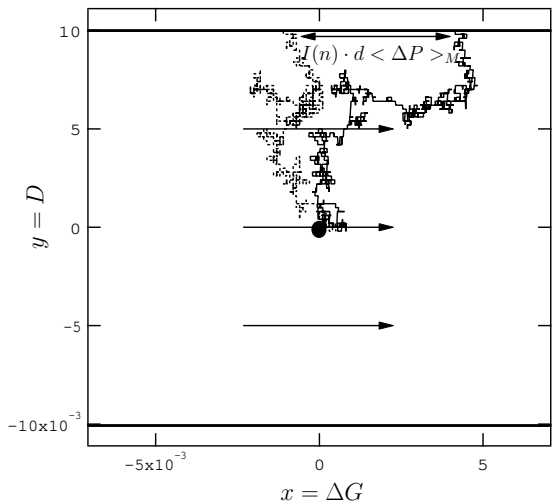}
  \end{center}
  \caption{Random walks on $\Delta G$ - $D$ space for Model-3. Two traces are plotted for comparison with the results of Model-1: The dotted line is for Model-1 and the solid line is for Model-3, both produced using the same random number generator. }
  \label{fig:FootPrint_D-DG_Model-3.eps}
\end{figure}

Adding this effect, the equations (\ref{eq:2D-brownian_M1}) in Model-1 are modified as
\begin{subequations}
\label{eq:2D-brownian_M3} 
\begin{eqnarray}
D(t+\Delta t)&=&D(t)+
\begin{cases}
+2c\Delta p\quad \text{(prob. 1/4)}\\
\pm 0\quad\quad\ \ \text{(prob. 1/2)}\ ,\\
-2c\Delta p\quad \text{(prob. 1/4)}
\end{cases}\label{eq:2D-brownian_M3-1}
\end{eqnarray}
\begin{eqnarray}
\Delta G(t+\Delta t)&=&\Delta G(t)+d<\Delta P>_M\Delta t\nonumber\\&&+
\begin{cases}
+c\Delta p\quad \text{(prob. 1/4)}\\
\pm 0\quad\quad\ \text{(prob. 1/2)}\ .\\
-c\Delta p\quad \text{(prob. 1/4)}
\end{cases}\label{eq:2D-brownianM3-2}
\end{eqnarray}
\end{subequations}
In the 2-dimensional random walk representation the initial conditions and the boundary conditions are invariant, however, we have a horizontal flow proportional to $d<\Delta P>_M$ as shown in Fig.\ref{fig:FootPrint_D-DG_Model-3.eps}. The existence of this flow implies that the distance of the absorption point from the origin is greater than for the original Model-1. As the vertical motions are completely identical the transaction interval is also identical; however, the absorbed point on the horizontal axis is shifted by $I(n)d<\Delta P>_M$. The strength of the flow depends on the parameter $d$ and the latest price changes.
 
\begin{figure}[htbp]
  \begin{center}
     \includegraphics[width=80mm]{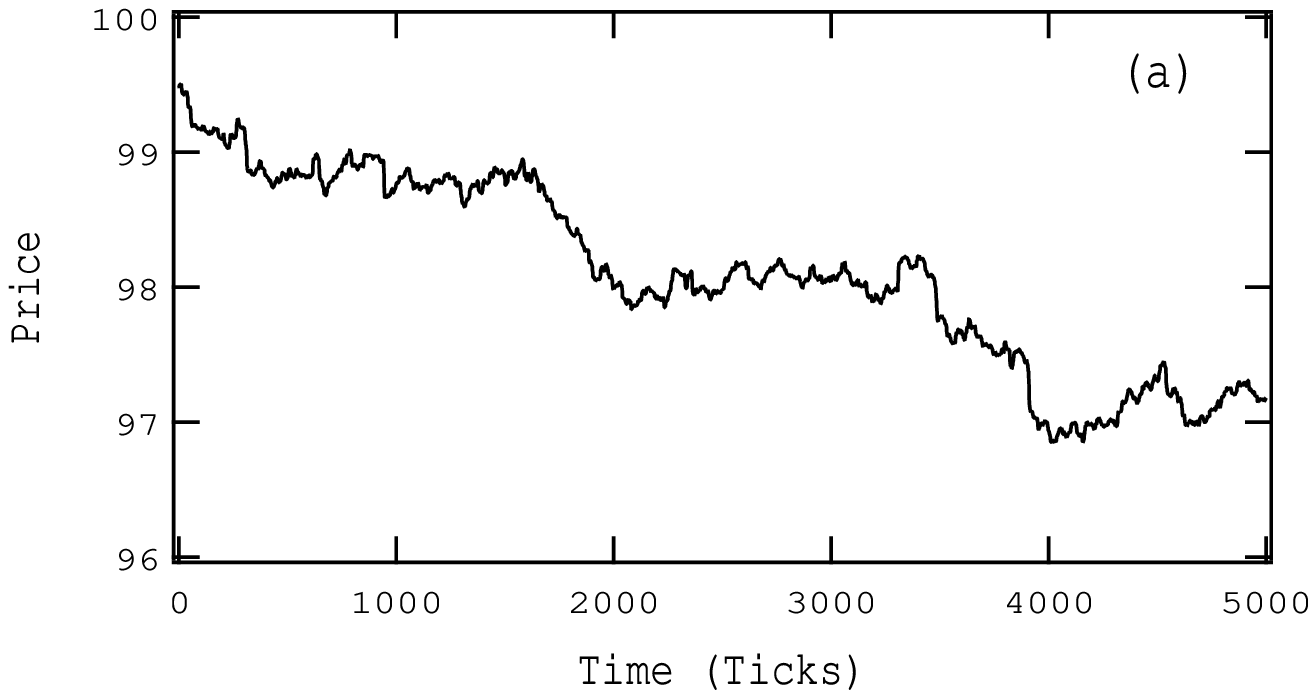}
    \includegraphics[width=70mm]{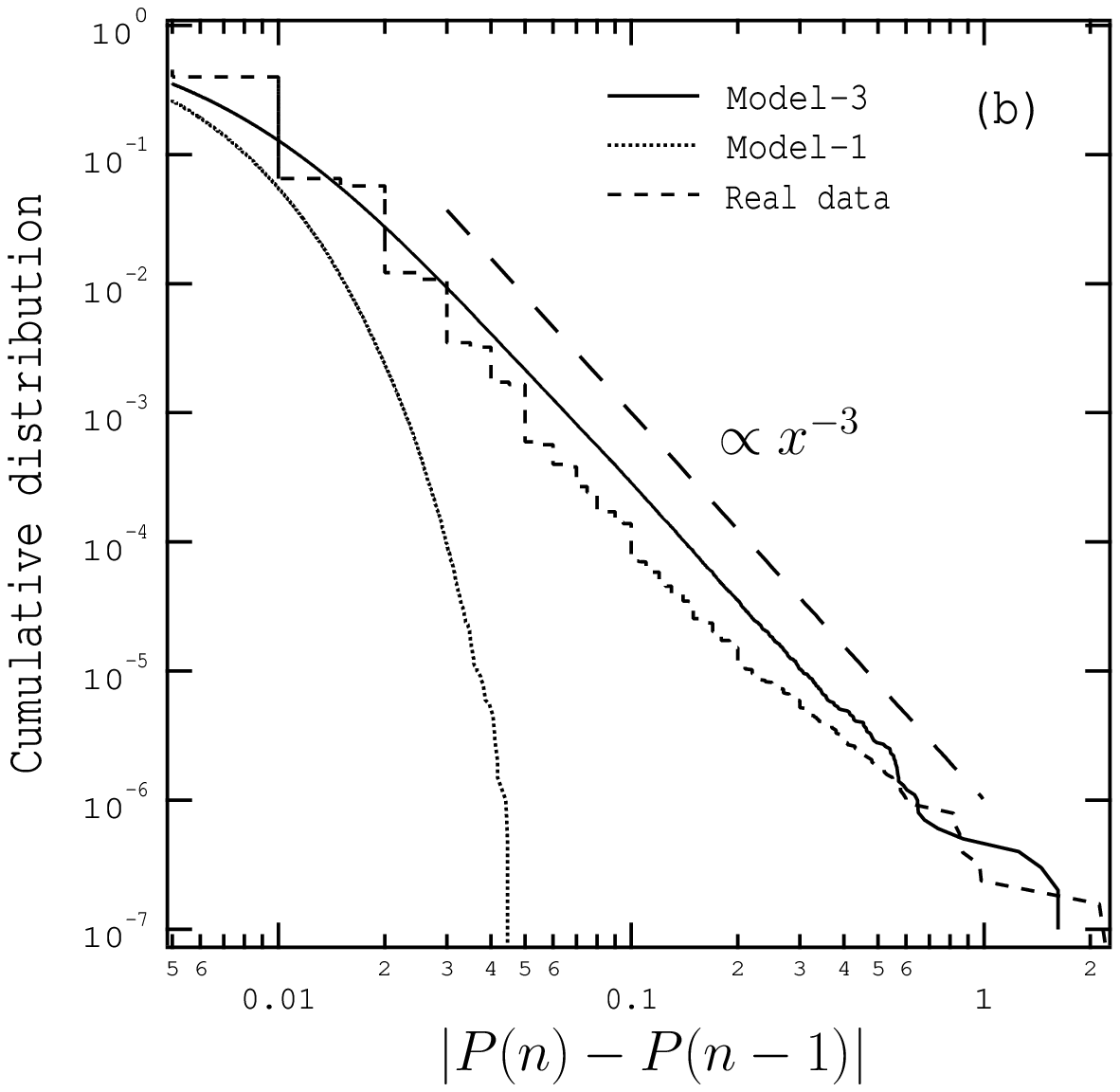}
  \end{center}
  \caption{Price changes produced by Model-3. (a) An example of time evolution of market prices produced by Model-3. (b) Cumulative distributions of price changes using a log-log scale. The dotted line is for Model-1; the solid line is for Model-3 and the dashed line for real data. The slope -3 is shown as a guide-line. }
\label{fig:CDF_ADP_Model3.eps}
\end{figure}

In this revised model the transaction intervals are identical to those of Model-1 because Eq.(\ref{eq:2D-brownian_M3-1}) is the same as Eq.(\ref{eq:2D-brownian_M1-1}), while the market price change is described by the following equation.
\begin{equation}
\Delta P(n+1)=I(n)\cdot d <\Delta P>_M+F(n).\label{eq:DP-model3}
\end{equation}
Here, the first term of the right hand side is the distance covered by the flow, and $d<\Delta P>_M$ gives the intensity of the flow, $I(n)$ is the transaction interval. The second term is identical to the price change of Model-1. From the results already obtained for Model-1 it is clear that both $I(n)$ and $F(n)$ are random variables characterized by exponential functions, so Eq.(\ref{eq:DP-model3}) follows a random multiplicative process. We know that a time series which is produced by a random multiplicative process generally follows a power law if the process satisfies a stationary condition. In particular, in the case that $M=1$ and $I(n)$ and $F(n)$ are independent in Eq.(\ref{eq:DP-model3}), we have an exact solution\cite{SMP1}. In this Model-3, $I(n)$ and $F(n)$ are not statistically independent and so we do not have an exact solution; however, if the effect of $F$ is negligibly small, then we can approximate the result using the solution for the independent case. We have the exponent $\beta$ for the power law of cumulative price change distribution as follows.
\begin{equation}
|d|^{\beta}<I(n)^{\beta}>=1.\label{eq:beta-1}
\end{equation}
Here $<I(n)^{\beta}>$ is the $\beta$-th order moment of $I(n)$, so we can apply Eq.(\ref{eq:moment-I}) to Eq.(\ref{eq:beta-1}). As a result, we have
\begin{equation}
|d|^{\beta}\frac{L^{2\beta}\Gamma(\beta+1)}{c^{2\beta}\Gamma(2\beta+1)}E_{\beta}=1.\label{eq:beta-2}
\end{equation}
It is known that empirical values of the power exponent of the cumulative price change distribution are around $-3$ in the actual market, so we set $\beta=3$, $L=0.01$ and $c=0.01$ in Eq.(\ref{eq:beta-2}).  Then we have $|d|\sim 1.25$. We can reproduce the power law of price change distribution with exponent $-3$ as represented in FIG.\ref{fig:CDF_ADP_Model3.eps}. Other parameters are taken to be $\Delta p=0.01$, $\Delta t=\Delta p\cdot \Delta p$ and $M=1$. 

\section{Relation Between the Dealer Model And the PUCK Model}

We have seen that the stochastic dealer models can reproduce important empirical features of markets. In this section we examine the relation to the market potential model called PUCK \cite{PUCK1}\cite{PUCK2}. In our previous work, we showed that the deterministic dealer model can reproduce market potentials confirmed by numerical simulation, and the essence for reconstruction of the potential is found to be the dealers' forecasting effect using moving averages\cite{Potential-DM}. We now present an analysis based on the present stochastic dealer model.

\begin{figure*}[htbp]
  \begin{center}
    \includegraphics[width=170mm]{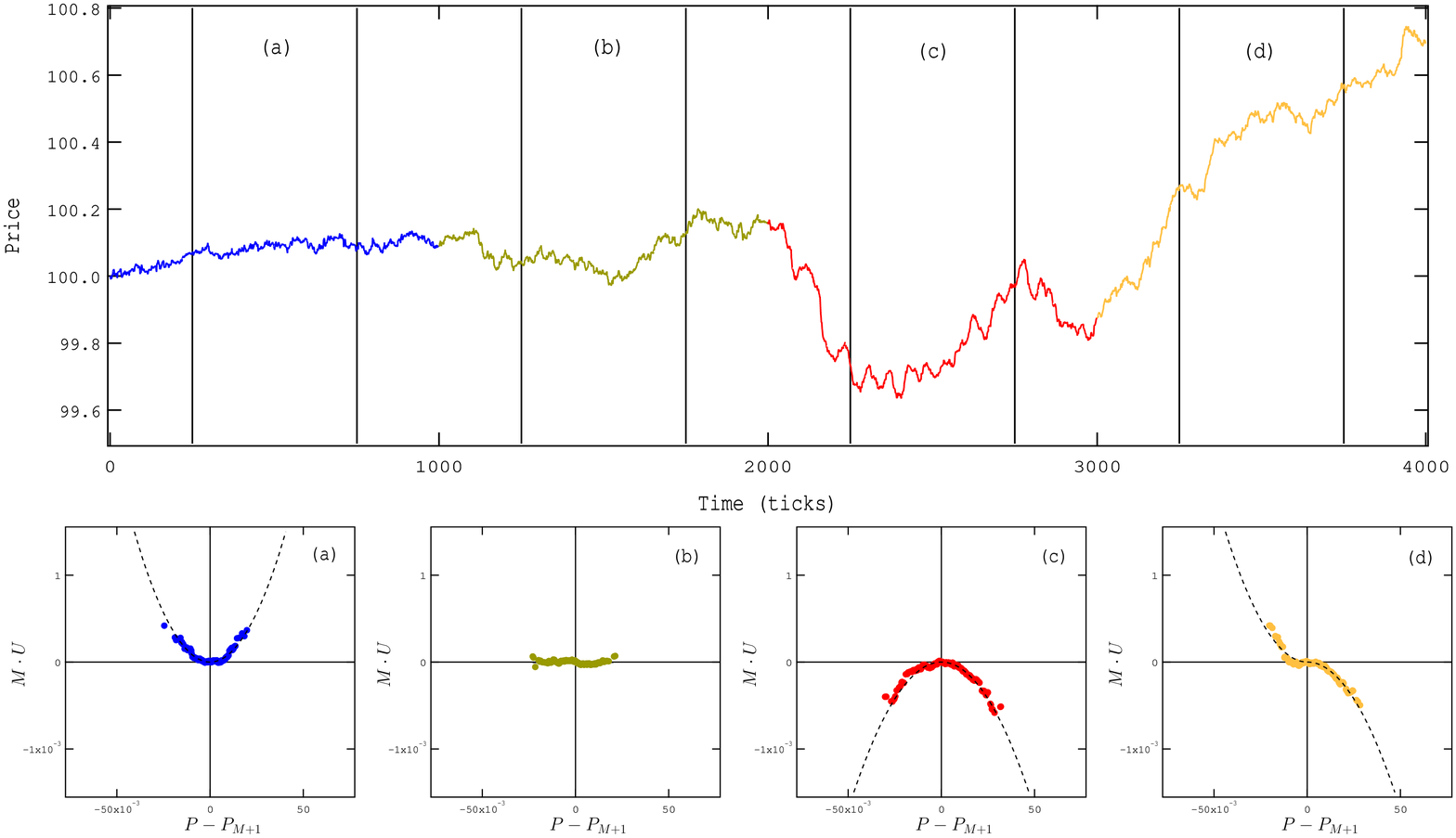}
  \end{center}
  \caption{Market potentials estimated for the time series produced by our stochastic dealer Model-3. The parameter values are changed: $d=-1.0$ from 1 to 1001 ticks, $d=0$ from 1001 to 2000 ticks, $d=1.0$ from 2001 to 3000 ticks, and from 3001 to 4000 ticks we set $d=1.0$ if $<\Delta P>_M\ge 0$ and $d=-1.0$ if $<\Delta P>_M<0$. Other parameters are given as follows: $c=0.01$, $L=0.01$, $\Delta p=0.01$, $\Delta t=\Delta p\cdot \Delta p$ and $M=10$. To estimate the market potential function, we use 500 ticks and $M=10$ and we show market potentials for the periods (a), (b), (c) and (d). The dashed line is approximated by a quadratic function, $y=ax^2$. Over the period (d), we fit the line in $x\ge 0$ and $x<0$ respectively by using quadratic functions as well as in (a), (b), (c).}
\label{fig:Potential_model3.eps}
\end{figure*}

It is easy to confirm that price changes produced by Model-3 yield non-trivial market potential functions as shown in FIG.\ref{fig:Potential_model3.eps}. In this figure the parameter $d$ in Eq.(\ref{eq:model3-1}) is changed. We set the parameter $d=-1.0$ during the period from n=1 to 1000 ticks, that is, the dealers' are contrarians who predict that the future price will move against the latest trend. The parameter $d=0$ in the period n=1001 to 2000 ticks, that is, the dealers are simple random walkers. And $d=1.0$ during 2001 to 3000 ticks, that is, the dealers are trend-followers predicting that near future prices will be proportional to a moving average of price changes. We can clearly observe a stable, a flat and an unstable potential function, respectively, as expected. During 3001 to 4000, we set $d=1.0$ when the average of the past $M$ price changes, $<\Delta P>_M$, is greater than or equal to zero, and we set $d=-1.0$ when $<\Delta P>_M<0$. In such an asymmetric case, we can find an asymmetric potential as shown in FIG.\ref{fig:Potential_model3.eps}d; in that case the market price increases nearly linearly on a large scale. It is apparent that the market potential function and the dealers' forecasting effect are also deeply related in this stochastic model.

The PUCK model is formulated as follows:
\begin{eqnarray}
P(n+1)&=&P(n)-\frac{\partial}{\partial P}U(P)\mid_{P=P(n)-P_{M+1}(n)}+F^{\prime}(n),\label{eq:PM-1}\nonumber\\
\\
U(P)&=&\frac{b(n)}{2M}P^2.\label{eq:PM-P}
\end{eqnarray}
Here $P(n)$ is the noise reduction price introduced by Ohnisi et. al., also referred to as the optimal moving average price\cite{OMA}, $U(P)$ is the potential function defined by Eq.(\ref{eq:PM-P}), $F^{\prime}(n)$ is an uncorrelated noise term, and $P_{M+1}(n)$ is the simple moving average over $M+1$ ticks: $\displaystyle{P_{M+1}(n)=\frac{1}{M+1}\sum_{k=0}^{M}P(n-k)}$. If the market potential is asymmetric as in FIG.\ref{fig:Potential_model3.eps}(d), we define the potential function for $x<0$ and $x\ge0$ respectively by using quadratic functions. In the case of symmetric potential as FIG.\ref{fig:Potential_model3.eps}abc, We can transform ($\ref{eq:PM-1}$) to the following equation:
\begin{equation}
\Delta P(n)=-\frac{b(n)}{2}<\Delta P>_{M}+F^{\prime}(n)\label{eq:PM-2}.
\end{equation}
$<\Delta P>_{M}$ is defined by Eq.($\ref{eq:WMA-1}$). We note that Eq.(\ref{eq:DP-model3}) in Model-3 and Eq.(\ref{eq:PM-2}) have the same form of a linear stochastic equation, so the statistical property is independent of the noise property. Comparing the coefficients of $<\Delta P>_{M}$ in Eq.(\ref{eq:DP-model3}) and Eq.(\ref{eq:PM-2}), we have the simple relation:
\begin{equation}
b(n)=-2d\cdot I(n).
\end{equation}
By taking the average over tick times $n$, we have 
\begin{equation}
<b>=-2d\cdot <I>=-d\left (\frac{L}{c}\right )^2,\label{eq: b-d}
\end{equation}
where $<x>$ denotes the average of $x$ over tick time $n$. This equation implies that the market is unstable ($b<0$) when dealers are trend followers ($d>0$) while the market is stable when dealers are contrarians ($d<0$). This result is consistent with our previous simulation results using the deterministic dealer model. 

\begin{figure}[htbp]
  \begin{center}
    \includegraphics[width=80mm]{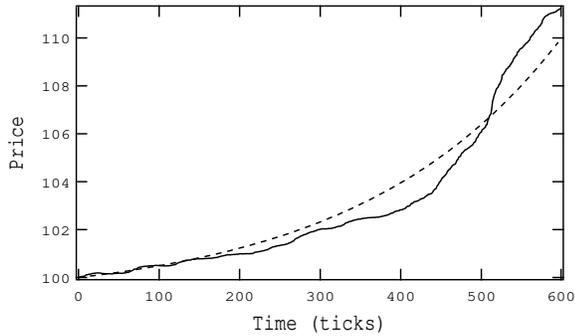}
  \end{center}
  \caption{A bubble-like phenomenon produced by Model-3. The dashed line is an exponential function of time as a guideline, $y=\exp(0.004x)+99$. The parameters satisfy $d\ge2c^2/L^2$; $d=2.0$, $c=0.01$ and $L=0.01$, $\Delta p=0.01$, $\Delta t=\Delta p\cdot \Delta p$ and $M=10$. }
\label{fig:Bubble_yahoo.eps}
\end{figure}

In the PUCK model, when the potential is of quadratic type such as in FIG.\ref{fig:Potential_model3.eps}abc, the diffusion coefficient of market prices, $\sigma$, is theoretically calculated as a function of the potential coefficient $b$\cite{PUCK-sigma}. By introducing the newly derived relation, Eq.(\ref{eq:sigma-b}), into the formula, we have a theoretical evaluation of the price diffusion coefficient for our Model-3. 
\begin{equation}
\sigma_d(\Delta n)=\frac{2c^2}{2c^2-dL^2}\sigma_{d=0}(\Delta n)\label{eq:sigma-b}.
\end{equation}
From this relation we find that in the range of $|d|\le2c^2/L^2$ the value of $\sigma$ is finite, and we can expect the market price to follow a normal random walk over a long time scale. When $d\ge2c^2/L^2$ the above formula is meaningless; however, the actual market price moves nearly monotonically or even exponentially as occurs for bubbles in real markets. Actually we can generate a bubble-like phenomenon by setting $d\ge2c^2/L^2$ as shown in  FIG.\ref{fig:Bubble_yahoo.eps}. Here, the time evolution is well approximated by an exponential function as predicted by the PUCK model\cite{PUCK-sigma}. We do not investigate this phenomenon further in this paper, but it should be stressed that our model can describe not only normal states of markets, but also abnormal states such as bubbles and crashes where prices move monotonically, by tuning our model's parameters.

\section{discussion}
We have introduced a new stochastic dealer model which consists of only two dealers, and showed that basic empirical laws of financial markets are well reflected in terms of transaction intervals and price changes. In Model-1, both dealers change their prices randomly, so fluctuations of transaction intervals and price changes are also random. We calculated these statistical properties exactly. The occurrence of transactions is well approximated by a Poisson process, and the price change distribution is well described by an exponential distribution.

In order to make our model more realistic the following two feedback effects were introduced. One was the feedback effect of transaction intervals, and the other was the feedback effect of price changes, both caused by the dealers' observations of the latest market status. As a result, the basic model's random noise properties were modulated and the artificial market reproduced both the distributions of transaction intervals and price changes to follow long-tailed distributions which are quite similar to those of real markets.

Moreover, we established that the dealers' action of prediction by using a moving average of past price changes generates market potentials in the PUCK formulation, and we derived a simple theoretical relation between the stochastic dealer model and the PUCK model. Namely, we found the relation between the microscopic dealers' strategy and the macroscopic market's stability as defined by the PUCK model.

As an application of the relation to the PUCK model, we checked the condition when market prices in our stochastic dealer model exhibit a bubble-like phenomenon in which price motion is approximated by an exponential function rather than a random walk. This transition from a random walk phase to an exponential growth phase is considered to be quite useful for discussions concerning how to realize a stable market.

We expect that our models can be used as a base for market experiments. For example, we may be able to observe the effect of governmental intervention by introducing a third dealer who only buys dollars over a given period. From the viewpoint of numerical simulation it is quite easy to increase the number of dealers, each of whom will have his own strategy. So we can construct any experimental market by adding or subtracting specific dealers, and observe the change in the market's macroscopic behavior as a result. For example, we may be able to find a way to avoid market crashes by introducing a specially designed dealer who acts to stabilize the market. This kind of market experiment may contribute to future attempts at real market stabilization.

\end{document}